\documentclass{PoS}

\usepackage{amsfonts,amssymb,graphicx}
\usepackage[tight]{subfigure}
\usepackage{rotating}
\usepackage[outercaption]{sidecap}
\usepackage[numbers,compress]{natbib}
\usepackage{longtable}
\usepackage{graphicx}
\usepackage{subfigure}
\usepackage[centertags]{amsmath}
\usepackage{multirow}

\newcommand{\bCentering}{\centering}
\newcommand{\bCaption}{\caption}

\newcommand{\unity}{{\footnotesize\mbox{1\!\!I}}}

\def\muc{\multicolumn}

\def\Z{\mathbb{Z}}
\def\unity{1\!\!{\rm I}}
\def\ov{\overline}

\def\Sym{\mathbf{Sym}}
\def\Anti{\mathbf{Anti}}

\def\ov{\overline}
\def\1{{\bf 1}}
\def\2{{\bf 2}}
\def\3{{\bf 3}}
\def\4{{\bf 4}}
\def\6{{\bf 6}}
\def\OR{\Omega\mathcal{R}}

\def\pp{\uparrow\uparrow}

\definecolor{mygr}{rgb}{0,0.6,0}
\definecolor{mygrey}{rgb}{0,0.1,0.2}
\definecolor{myblue}{rgb}{0,0.5,0.9}
\definecolor{myblue2}{rgb}{0,0.5,0.5}
\definecolor{myorange}{rgb}{1,0.5,0}
\definecolor{mypurple}{rgb}{0.6,0,1}
\definecolor{mygolden}{rgb}{1,0.8,0.2}

\newcommand{\bCaptionfonts}{\small}
\makeatletter
\long\def\@makecaption#1#2{%
  \vskip\abovecaptionskip
  \sbox\@tempboxa{{\bCaptionfonts #1: #2}}%
  \ifdim \wd\@tempboxa >\hsize
    {\bCaptionfonts #1: #2\par}
  \else
    \hbox to\hsize{\hfil\box\@tempboxa\hfil}%
  \fi
  \vskip\belowcaptionskip}
\makeatother

\makeatletter
\let\ORIGINALlatex@openbib@code=\@openbib@code
\renewcommand{\@openbib@code}{\ORIGINALlatex@openbib@code\setlength{\itemsep}{1ex plus.5ex minus.5ex}\setlength{\parsep}{0pt}}
\makeatother

\def\mathtabfix#1#2#3{\begin{table}[th]\bCentering\resizebox{\linewidth}{!}{$#1$}\bCaption{#3}\label{tab:#2}\end{table}}

\title{D6-Brane Model Building and Discrete Symmetries on $T^6/(\Z_2\times \Z_6' \times \OR)$ with Discrete Torsion}

\ShortTitle{D6-Brane Model Building and Discrete Symmetries on $T^6/(\Z_2\times \Z_6' \times \OR)$}

\author{Gabriele Honecker and \speaker{Wieland Staessens}\thanks{The work of both authors is partially supported by the {\it Cluster of Excellence `Precision Physics, Fundamental Interactions and Structure of Matter' (PRISMA)} DGF no. EXC 1098,
the DFG research grant HO 4166/2-1 and the Research Center {\it `Elementary Forces and Mathematical Foundations' (EMG)} at JGU Mainz.
}\\
      PRISMA Cluster of Excellence \&
Institut f\"ur Physik  (WA THEP), Johannes-Gutenberg-Universit\"at, D-55099 Mainz, Germany\\
        E-mail: \email{Gabriele.Honecker@uni-mainz.de},  \email{Wieland.Staessens@uni-mainz.de}}

\abstract{We review several geometric aspects and properties of the orbifold $T^6/(\Z_2\times \Z_6' \times \OR)$ with discrete torsion, that are crucial with respect to global model building and the search for discrete gauge symmetries in the context of intersecting D6-brane models. A global six-stack Pati-Salam model is used for illustration, and various characteristics of its effective field theory are discussed.}

\FullConference{Proceedings of the Corfu Summer Institute 2012 "School and Workshops on Elementary Particle Physics and Gravity"\\
		September 8-27, 2012\\
		Corfu, Greece
		\hspace{4.2in}
		{\rm MITP/13-021}}

\begin{document}

\section{Introduction}\label{S:Intro}
Intersecting D6-brane models on Calabi-Yau orientifolds offer a nice geometric picture, in which the chiral spectrum and gauge interactions of the Standard Model consistently coexist with quantum gravity and yield an ${\cal N} = 1$ supersymmetric effective four-dimensional theory, see e.g.~\cite{Blumenhagen:2006ci}. While stacks of identical D6-branes provide for the gauge interactions, matter arises as chiral multiplets at the intersection points of the three-cycles wrapped by the D6-branes along the six compact dimensions. Toroidal orbifolds offer excellent backgrounds to construct 
phenomenologically appealing consistent models, see e.g.~\cite{Cvetic:2001tj}, with a well-developed understanding of the perturbative characteristics of the effective four-dimensional field theory, see e.g.~\cite{Cremades:2003qj}.

For the simplest toroidal orbifolds, the D6-brane stacks come automatically with matter multiplets in the adjoint representation of the $SU(N)$ gauge group, encompassing the position and Wilson line moduli of the D6-branes along the compact directions. The more intricate toroidal orbifold $T^6/(\Z_2 \times \Z_6')$ with discrete torsion contains a $\Z_2\times \Z_2$ subsymmetry allowing to project out this type of adjoint matter completely. As the D6-branes are stuck at the $\Z_2\times \Z_2$ fixed points, the gauge group can no longer be spontaneously broken by displacing the D6-branes continuously along one of compact directions~\cite{Blumenhagen:2005tn,Forste:2010gw}. The appearance of so-called `rigid' three-cycles entices to search for globally defined D6-brane models on $T^6/(\Z_2 \times \Z_6')$ with discrete torsion providing for a chiral spectrum and gauge interactions as in the MSSM or GUTs~\cite{Honecker:2012qr}. 

In the MSSM discrete symmetries, such as R-parity, are invoked in order to forbid couplings leading to a large proton decay rate~\cite{Ibanez:1991pr}. In the context of intersecting D6-brane models such discrete $\Z_n$ symmetries are of stringy origin~\cite{BerasaluceGonzalez:2011wy,Ibanez:2012wg,Honecker:2013hda} , as they arise from $U(1)$ gauge factors acquiring a string scale St\"uckelberg mass by virtue of the generalized Green-Schwarz-mechanism. Interestingly, discrete $\Z_n$ symmetries with a gauge origin are left unbroken by non-perturbative effects, like D-brane instantons~\cite{Blumenhagen:2009qh}, and therefore present perfect elements to constrain the non-perturbative sector of the theory.

In these proceedings, we review the geometric properties of the orientifold $T^6/(\Z_2 \times \Z_6'\times \OR)$ with discrete torsion and briefly discuss the pairwise identification of the a priori four different lattice configurations, by which the number of backgrounds to construct global models with large hidden gauge groups is significantly reduced. We summarize the procedure to search for discrete $\Z_n$ symmetries on this orbifold, for which the classification of rigid three-cycles with an enhanced $USp(2N)$ or $SO(2N)$ gauge group is required. These formal aspects of the orbifold are clarified by an elaborate analysis of an explicit global six-stack Pati-Salam model.

\section{Geometric properties of $T^6/(\Z_2\times \Z_6' \times \OR)$}\label{S:Geometry}
The orbifold $T^6/(\Z_2\times \Z_6' )$ is constructed as a factorisable six-torus, for which the complex coordinates $z^k$ of each two-torus $T^{2}_{(k=1,2,3)}$ are identified under the $\Z_2\times \Z_6'$ action:
\begin{equation}\label{Eq:Z2Z6p-def}
\theta^m \, \omega^n: \; z^k \to e^{2\pi \, i \, (m v_k + n w_k^{\, \prime})} \; z^k
\qquad
\text{with}
\qquad
\vec{v}= \frac{1}{2}(1,-1,0),
\qquad
\vec{w}^{\, \prime} = \frac{1}{6}(-2,1,1).
\end{equation}
The cristallographic action of the $\Z_6'$ generator $\omega$ freezes the complex structure modulus for each two-torus, forcing each two-torus to take the shape of an $SU(3)$ root lattice. The orbifold group $\Z_2 \times \Z_6'$ contains one $\Z_3$ subgroup and three equivalent $\Z_2^{(k)}$ subgroups with 16 fixed points each along the four-tori $T^2_{(i)}\times T^2_{(j)} \equiv T^4_{(k)}$ and the invariant two-torus $T^{2}_{(k)}$ under $\Z_2^{(k)}$, with $(k,i,j)$ a cyclic permutation of $(1,2,3)$. Furthermore, the generator of $\Z_2$ can act with a phase factor $\eta = \pm 1$ on the $\Z_6'$ twisted sector and vice versa. The non-trivial choice $\eta = - 1$, so called `with discrete torsion', will be assumed for the remainder of the article.

\begin{SCfigure}[50][h]
\includegraphics[width=0.3\textwidth]{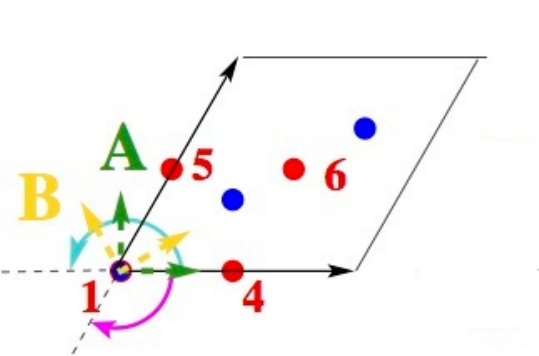} \begin{picture}(0,0) \put(-45,10){$\pi_{2i-1}$} \put(-95,70){$\pi_{2i}$} \put(-50,75){$r_i$} \end{picture}
 \caption{$SU(3)$ lattice for a two-torus $T^2_{(i),i\in\{1,2,3\}}$ in the factorisable $T^6/(\Z_2\times \Z_6' )$ orbifold: $\Z_3$ fixed points are indicated in blue, $\Z_2$ fixed points \{1,4,5,6\} in red. The symmetry axis of the anti-holomorphic involution $\cal R$ can lie along the one-cycle $\pi_{2i-1}$ ({\bf A} lattice) or along the one-cycle $\pi_{2i-1} + \pi_{2i}$ ({\bf B} lattice).  \label{T6Z2Z6p}}
\end{SCfigure}
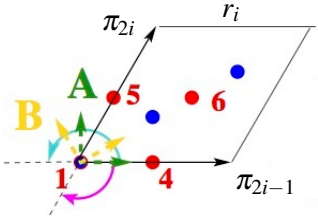

For model building purposes in Type IIA string theory, the point group of the orbifold has to be accompanied by an orientifold projection $\OR$ consisting of the worldsheet parity $\Omega$ and an anti-holomorphic involution ${\cal R}: z^k \rightarrow \bar z^k$. The hexagonal shape of the two-torus lattice only allows two possible orientations ({\bf A} and {\bf B}) for the symmetry axis of $\cal R$, as depicted in figure~\ref{T6Z2Z6p}. The geometric indistinguishability of the three two-tori then leads to four inequivalent lattice configurations: {\bf AAA}, {\bf AAB}, {\bf ABB} and {\bf BBB}~\cite{Forste:2010gw}.

For each lattice configuration, the orientifold O6-planes fall into four different orbits under the $\Z_6'$ action, labeled as $\OR$ and 
$\OR\Z_2^{(k),k\in\{1,2,3\}}$, respectively. Worldsheet consistency links the discrete torsion phase $\eta$ to the sign of the RR-charges of the O6-planes ($\eta_{\OR}$, $\eta_{\OR \Z_2^{(k)}}$):
\begin{equation}
\eta= \eta_{\OR} \prod_{k=1}^3 \eta_{\OR\Z_2^{(k)}}
,
\end{equation}
implying that one of the O6-planes has to be an exotic O6-plane with negative RR charge for orbifolds with discrete torsion ($\eta=-1$).

\begin{SCfigure}[50][h]
\vspace{0.3in}\includegraphics[width=0.25\textwidth]{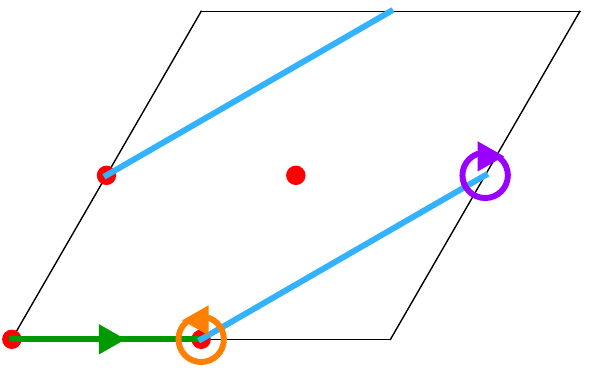}  \begin{picture}(0,0) \put(-35,0){$\pi_{2i-1}$} \put(-90,70){$\pi_{2i}$} \put(-95,-10){\color{mygr} $\sigma_a^i$} \put(-80,15){\color{myorange} $\tau_a^{\Z_2^{(k)}}$} \put(-20,25){\color{mypurple} $\tau_a^i$} \end{picture}
\caption{Discrete parameters describing the location of a fractional cycle on a two-torus $T^{2}_{(i)}$: the displacement $\sigma_a^i \in \{0,1\}$ indicates whether the bulk cycle passes through the origin or is shifted from it by half a lattice vector; the parameter $\tau_a^{\Z_2^{(k)}} \in \{0,1 \}$ can be viewed as whether the exceptional part of the fractional cycle encircles the first fixed point clockwise or counter-clockwise; the discrete Wilson line $\tau_a^i \in \{0,1 \}$ then indicates whether the fractional cycle encircles the second fixed point with the same orientation or the opposite one as the first fixed point.} \label{T2ExcepDiscreteParameters}
\end{SCfigure}
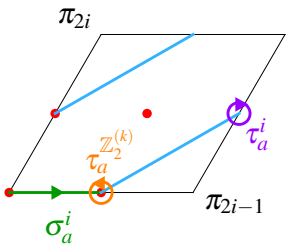

Supersymmetric D6-branes on this background fill the four-dimensional Minkowski spacetime and wrap fractional three-cycles satisfying the special Langrangian conditions. The fractional three-cycles are constructed as linear combinations of bulk and exceptional three-cycles at $\Z_2^{(i)}$ fixed points: 
\begin{equation}
\Pi^{\text{frac}}_a \equiv  \frac{1}{4} \left(\Pi^{\text{bulk}}_a + \sum_{i=1}^3 \Pi^{\Z_2^{(i)}}_a  \right).
\end{equation}
The bulk three-cycles are inherited from the underlying factorisable six-torus and are decomposable in terms of $b_3^{\text{bulk}}=2+2 \, h_{21}^U=2$ orbifold-invariant factorisable three-cycles:
\begin{equation}
\rho_1 = 4 \left( 1 + \omega + \omega^2 \right) \pi_{135}, \qquad
\rho_2 = 4 \left( 1 + \omega + \omega^2 \right) \pi_{136} 
,
\end{equation}
with intersection number 
\begin{equation}\label{Eq:IntersectionBulk}
 \rho_1 \circ \rho_2  =4 .
 \end{equation}
Hence a bulk three-cycle is expressed as 
\begin{equation}
\Pi^{\text{bulk}}_a = X_a\, \rho_1 + Y_a\, \rho_2, 
\end{equation}
with the bulk wrapping numbers $(X_a, Y_a)$ defined in terms of the toroidal wrapping numbers $(n_a^i,m_a^i)_{i=1,2,3}$:
\begin{equation}\label{Eq:Z2Z6p-Def-XY}
\begin{aligned}
X_a \equiv  n^1_a n^2_a n^3_a  - m^1_a m^2_a m^3_a -\sum_{i\neq j\neq k\neq i} n^i_a m^j_a m^k_a
,
\qquad
Y_a \equiv \!\! \sum_{i\neq j\neq k\neq i} \!\! \left(n^i_a n^j_a m^k_a + n^i_a m^j_a m^k_a  \right)
.
\end{aligned}
\end{equation} 
An exceptional three-cycle  corresponds to the orbifold-invariant product of an exceptional divisor $e^{(k)}_{xy}$ stuck at the $\Z^{(k)}_2$ fixed points
$xy \in T^{2}_{(i)} \times T^2_{(j)} \equiv T^4_{(k)}$ (with $x,y \in \{ 1, 4, 5, 6 \}$)
 with a toroidal one-cycle along the invariant two-torus~$T^2_{(k)}$. For each $\Z_2^{(k)}$ twisted sector, a basis of ten $\Z_6'$-invariant exceptional three-cycles $(\varepsilon^{(k)}_\alpha, \tilde \varepsilon^{(k)}_\alpha)_{\alpha \in \{ 1, \ldots, 5 \}}$ is introduced:
\begin{equation}
\epsilon_\alpha^{(k)} = 2 (1 + \omega + \omega^2) [ e_{xy}^{(k)} \otimes \pi_{2k-1}  ] , \qquad \tilde \epsilon_\alpha^{(k)} = 2 (1 + \omega + \omega^2) [ e_{xy}^{(k)} \otimes \pi_{2k}  ].
\end{equation}
The intersection form for the orbifold-invariant exceptional three-cycles reads
\begin{equation}\label{Eq:IntersectionExceptional}
\varepsilon^{(k)}_{\alpha} \circ\tilde{\varepsilon}^{(l)}_{\beta} = -4 \; \delta^{kl} \, \delta_{\alpha\beta}
.
\end{equation}
A generic exceptional three-cycle $\Pi_a^{\Z_2^{(k)}}$ in the $\Z_2^{(k)}$ twisted sector can then be expressed in terms of the exceptional three-cycle basis as
\begin{equation}
\Pi_a^{\Z_2^{(k)}}=\sum_{\alpha=1}^5 \left[x^{(k)}_{\alpha,a} \; \varepsilon^{(k)}_{\alpha}
+  y^{(k)}_{\alpha,a} \; \tilde{\varepsilon}^{(k)}_{\alpha} \right],
\end{equation}
where the exceptional wrapping numbers $(x^{(k)}_{\alpha,a}, y^{(k)}_{\alpha,a})$ depend on the torus wrapping numbers, the $\Z_2^{(k)}$ eigenvalue\footnote{Although each $\Z_2^{(k)}$ twisted sector comes with its own eigenvalue, the three discrete parameters $\tau_a^{\Z_2^{(k)}}$ are not completely independent, but satisfy one relation: $(-1)^{\tau_a^{\Z_2^{(1)}}+\tau_a^{\Z_2^{(2)}}}=(-1)^{\tau_a^{\Z_2^{(3)}}}$. As the sequence of a $\Z_2^{(1)}$ action followed by a $\Z_2^{(2)}$ action on the complex coordinates boils down to a $\Z_2^{(3)}$ action, the discrete eigenvalues are bound to satisfy this constraint.} $(-1)^{\tau_a^{\Z_2^{(k)}}}$, the discrete displacements $(\sigma_{a}^i, \sigma_{a}^j)$ and the discrete Wilson lines $(\tau_a^i, \tau_a^j)$ along the four-torus $T^2_{(i)}\times T_{(j)}^2$. Figure~\ref{T2ExcepDiscreteParameters} schematically displays the geometric meaning of the discrete parameters. The construction of an exceptional three-cycle $\Pi_a^{\Z_2^{(k)}}$ for a given bulk three-cycle is done by using table~\ref{tab:Z2Z6p-fps+excycles}, once it is clear from the toroidal wrapping numbers $(n^i_a,m^i_a)$ and displacements $(\sigma_{a}^i, \sigma_{a}^j)$ which fixed points are traversed by the bulk three-cycle. More details about the construction procedure of supersymmetric fractional three-cycles are given in~\cite{Forste:2010gw} and~\cite{Honecker:2012qr}.

\mathtabfix{
\begin{array}{|c|c||c|c|}\hline
\multicolumn{4}{|c|}{\Z_2^{(k)} \; \text{\bf fixed points and exceptional 3-cycles on } \, T^6/\Z_2 \times \Z_6' \; \text{\bf with discrete torsion } (\eta=-1)} 
\\\hline\hline
{\rm f.p.}^{(k)} \otimes (n^{k}_a \pi_{2k-1} + m^{k}_a \pi_{2k}) & {\rm orbit}  & {\rm f.p.}^{(k)} \otimes (n^{k}_a \pi_{2k-1} + m^{k}_a \pi_{2k}) & {\rm orbit} 
\\\hline\hline
11 & - 
& 55 &  -(n^k_a + m^k_a)  \varepsilon^{(k)}_3 + n^k_a \tilde{\varepsilon}^{(k)}_3
\\\cline{1-2}
41 &   n^k_a \; \varepsilon^{(k)}_1 + m^k_a \tilde{\varepsilon}^{(k)}_1
&66 &  m^k_a \varepsilon^{(k)}_3 - (n^k_a + m^k_a)  \tilde{\varepsilon}^{(k)}_3
\\\cline{3-4}
51 & -(n^k_a + m^k_a)  \varepsilon^{(k)}_1 + n^k_a \tilde{\varepsilon}^{(k)}_1
& 45 &    n^k_a \varepsilon^{(k)}_4 + m^k_a \tilde{\varepsilon}^{(k)}_4
\\
61 & m^k_a \varepsilon^{(k)}_1 - (n^k_a + m^k_a)  \tilde{\varepsilon}^{(k)}_1
& 56 &  -(n^k_a + m^k_a)  \varepsilon^{(k)}_4 + n^k_a \tilde{\varepsilon}^{(k)}_4
\\\cline{1-2}
14 &   n^k_a \varepsilon^{(k)}_2 + m^k_a \tilde{\varepsilon}^{(k)}_2
& 64 &  m^k_a \varepsilon^{(k)}_4 - (n^k_a + m^k_a)  \tilde{\varepsilon}^{(k)}_4
\\\cline{3-4}
15 & -(n^k_a + m^k_a)  \varepsilon^{(k)}_2 + n^k_a \tilde{\varepsilon}^{(k)}_2
& 46 &    n^k_a \varepsilon^{(k)}_5 + m^k_a \tilde{\varepsilon}^{(k)}_5
\\
16 &  m^k_a \varepsilon^{(k)}_2 - (n^k_a + m^k_a)  \tilde{\varepsilon}^{(k)}_2
& 54 &  -(n^k_a + m^k_a)  \varepsilon^{(k)}_5 + n^k_a \tilde{\varepsilon}^{(k)}_5
\\\cline{1-2}
44 &   n^k_a \varepsilon^{(k)}_3 + m^k_a \tilde{\varepsilon}^{(k)}_3
& 65 &  m^k_a \varepsilon^{(k)}_5 - (n^k_a + m^k_a)  \tilde{\varepsilon}^{(k)}_5
\\\hline
\end{array}
}{Z2Z6p-fps+excycles}{Complete list of orbits of exceptional divisors $e_{xy}^{(k)}$ at fixed points $xy$ of $\Z_2^{(k)}$ along $T^4_{(k)}$ times one-cycles $n^{k}_a \pi_{2k-1} + m^{k}_a \pi_{2k}$ 
along $T^2_{(k)}$ under  the $\Z_6'$ orbifold generator $\omega$ as computed in~\cite{Forste:2010gw}. In order to construct the full orbifold-invariant exceptional three-cycles, the orbits have to be dressed with the appropriate $\Z_2^{(k)}$ eigenvalue and the discrete Wilson lines $(\tau_a^i, \tau_a^j)$.} 
\begin{table}[h]
\hspace*{-0.6in}
\begin{tabular}{|@{\hspace{0.in}}c@{\hspace{-0.1in}}c@{\hspace{0.in}}c||@{\hspace{0.in}}c@{\hspace{-0.1in}}c@{\hspace{-0.in}}c|}
\hline \multicolumn{6}{|c|}{\bf Pairwise identification of the lattices on $T^6/(\Z_2\times \Z_6' \times \OR)$ with $\eta=-1$}\\
\hline
\hline
 {\bf AAA} & $ \longleftrightarrow$ & \hspace{-0.8in} {\bf ABB} & {\bf AAB} &$ \longleftrightarrow$& \hspace{-0.8in}{\bf BBB} \\
\hline \hline
\rule{0pt}{4ex}  $\left(\begin{array}{cc} n^1_a & m^1_a  \\ n^2_a  & m^2_a  \\ n^3_a  & m^3_a   \end{array}\right)$ & $ =$ &\hspace{-0.6in} $ \left(\begin{array}{cc} \ov{n}^1_a  + \ov{m}^1_a  & - \ov{n}^1_a  \\ \ov{n}^2_a  & \ov{m}^2_a  \\ \ov{n}^3_a  & \ov{m}^3_a   \end{array}\right)$ 
 &  $\left(\begin{array}{cc} n^1_a & m^1_a  \\ n^2_a  & m^2_a  \\ n^3_a  & m^3_a   \end{array}\right)
$&$=$& \hspace{-0.6in}  $\left(\begin{array}{cc} \ov{n}^1_a  & \ov{m}^1_a  \\ \ov{n}^2_a  & \ov{m}^2_a  \\ \ov{n}^3_a +\ov{m}^3_a   & -\ov{n}^3_a   \end{array}\right) $ \\
\rule{0pt}{4ex} 
 $ \left(X_a,Y_a \right)$ & $=$ &\hspace{-0.6in}  $  \left(\ov{X}_a+\ov{Y}_a, - \ov{X}_a \right) $ & $\left(X_a,Y_a \right)$ & $=$ &\hspace{-0.6in}   $ \left(\ov{X}_a+\ov{Y}_a, - \ov{X}_a \right)$\\
 \rule{0pt}{4ex}   
$ (\OR, \OR\Z_2^{(1)}) $ & $=$&\hspace{-0.8in} $(\ov{\OR\Z_2^{(1)}}, \ov{\OR})$ &$ (\OR, \OR\Z_2^{(3)}) $ & $=$&\hspace{-0.8in} $(\ov{\OR\Z_2^{(3)}}, \ov{\OR})$ \\

 $(\OR\Z_2^{(2)},\OR\Z_2^{(3)})$ & $=$& \hspace{-0.8in}  $(\ov{\OR\Z_2^{(3)}},\ov{\OR\Z_2^{(2)}})$ & $(\OR\Z_2^{(1)},\OR\Z_2^{(2)})$ & $=$& \hspace{-0.8in}  $(\ov{\OR\Z_2^{(2)}},\ov{\OR\Z_2^{(1)}})$\\
 \rule{0pt}{4ex}   
 $\left( x_{\alpha,a}^{(1)} \, , \,  y_{\alpha,a}^{(1)} \right)_{\alpha \in \{1 \ldots 5\}}$ & $=$& $\left( \ov{x}_{\alpha,a}^{(1)} + \ov{y}_{\alpha,a}^{(1)} \, , \, -\ov{x}_{\alpha,a}^{(1)} \right)_{\alpha \in \{1 \ldots 5\}}$ 
 & $\left(x^{(3)}_{\alpha,a} \, ,\, y^{(3)}_{\alpha,a} \right)_{\alpha \in \{1 \ldots 5\}}$ & $=$ & $\left( \ov{x}^{(3)}_{\alpha,a} + \ov{y}^{(3)}_{\alpha,a} \, , \,  - \ov{x}^{(3)}_{\alpha,a} \right)_{\alpha \in \{1 \ldots 5\}}$  \\
 \rule{0pt}{4ex}   
 $\left( x^{(j)}_{\alpha,a} \, , \, y^{(j)}_{\alpha,a} \right)^{j=2,3}_{ \alpha=(1,3,4,5)}$ & $=$ &  $\left( \ov{y}^{(j)}_{\ov{\alpha},a} \, , \, - ( \ov{x}^{(j)}_{\ov{\alpha},a} + \ov{y}^{(j)}_{\ov{\alpha},a} ) \right)^{j=2,3}_{\ov{\alpha}=(1,5,3,4)}$ 
 & $\left(x^{(i)}_{\alpha,a}\, , \, y^{(i)}_{\alpha,a} \right)^{i =1,2}_{ \alpha = (1,3,4,5)}$& $=$& $\left( - (\ov{x}^{(i)}_{\ov \alpha,a} + \ov{y}^{(i)}_{\ov \alpha,a}) \, , \, \ov{x}^{(i)}_{\ov \alpha,a} \right)^{i =1,2}_ {\ov{\alpha} = (1,5,3,4)}$\\
 \rule{0pt}{4ex}   
 $\left( x^{(j)}_{2,a} \, , \, y^{(j)}_{2,a} \right)^{j=2,3}$ & $=$ & \hspace{-0.6in} $\left( \ov{x}^{(j)}_{2,a} \, , \, \ov{y}^{(j)}_{2,a} \right)^{j=2,3}$ & $\left( x^{(i)}_{2,a}\, ,\, y^{(i)}_{2,a} \right)^{i =1,2}$&$=$& $\left( \ov{y}^{(i)}_{2,a}, - ( \ov{x}^{(i)}_{2,a} +\ov{y}^{(i)}_{2,a} ) \right)^{i =1,2}$\\
 \hline 
\end{tabular} \begin{picture}(0,0) \put(-50,165){ \begin{rotate}{90} \it torus  \end{rotate}} \put(-47,130){\begin{rotate}{90}\it bulk \end{rotate}} \put(-47,90){\begin{rotate}{90} \it O6-pl. \end{rotate}} \put(-47,15){\begin{rotate}{90}\it exceptional \end{rotate}} \end{picture}
\begin{center}
\caption{Pairwise identification of the four lattice configurations for $T^6/(\Z_2\times \Z_6' \times \OR)$ with discrete torsion. The first row denotes the identification of the torus wrapping numbers under the non-supersymmetric rotation by $\pm \pi/3$ along $T_{(1)}^2$ for the {\bf AAA}-{\bf ABB} equivalence, and along $T^2_{(3)}$ for the {\bf AAB}-{\bf BBB}  equivalence. The second row lists the identification of the bulk wrapping numbers, and the subsequent two rows give the identification of the O6-planes. The last three rows express the identification of the exceptional wrapping numbers per twisted sector. The discrete parameters ($\Z_2^{(k)}$ eigenvalues, displacements $\sigma^k$ , Wilson lines $\tau^k$) of a three-cycle remain unchanged under the equivalence relations.}\label{Tab:PairwiseIdent}
\end{center}
\end{table}

In~\cite{Honecker:2012qr} it was shown that the four a priori distinct lattices can be identified two-by-two by performing a non-supersymmetric rotation on the lattices: the {\bf ABB} lattice is identified with the {\bf AAA} lattice by a rotation of the basis one-cycles over $\pi/3$ along the first two-torus, and the {\bf BBB} lattice with the {\bf AAB} lattice by a rotation over $\pi/3$ along the third two-torus. The full pairwise identification for the toroidal, bulk and exceptional wrapping numbers is given in table~\ref{Tab:PairwiseIdent}. As the identification respects the lengths of the three-cycles, also the O6-planes have to be permuted. Under this identification, the massless open and closed string spectra are fully equivalent, and the equivalence can also be proven at the level of the global consistency conditions (RR tadpole cancellation, K-theory constraint and supersymmetry), intersection numbers and string one-loop amplitudes without operator insertions, as shown in full length in~\cite{Honecker:2012qr}. The identification  significantly simplifies
D6-brane model building on this orbifold, as only two inequivalent lattices remain: {\bf AAA} and {\bf BBB}, where only the first lattice configuration turns out to be a promising background for model building. More explicitly, the {\bf BBB} lattice configuration does not allow for the construction of global models with large hidden gauge groups, as the (bulk) RR charges of the O6-planes are for the most part already canceled by the RR charges of the Standard Model D6-branes. 
Henceforth, only the {\bf AAA} lattice configuration will be considered and discussed. The bulk RR tadpole cancellation conditions for the {\bf AAA} lattice read:
\begin{equation}
\sum_a N_a \left(2 \,X_a+Y_a \right)= 4 \left( \eta_{\OR} + 3 \, \sum_{i=1}^3 \eta_{\OR\Z_2^{(i)}} \right). 
\end{equation}
Global models with a large hidden gauge group can only be found if the $\OR$-plane is taken to be the exotic O6-plane ($\eta_{\OR}=-1$), for which the net RR-charge of the O6-planes is maximal. The special Lagrangian conditions translate into two explicit conditions for the bulk part of the fractional cycle, 
\begin{equation}
Y_a=0, \hspace{0.1in} \text { and }  \hspace{0.1in} 2 \, X_a + Y_a > 0,
\end{equation}
which ensure that the D6-branes are calibrated w.r.t.~the same holomorphic three-form as the O6-planes.

In deriving the bulk RR tadpole cancellation conditions, the behaviour of the bulk cycles under the orientifold projection has implicitly been used. The $\OR$ projection acting on an {\bf A} lattice leaves the basic one-cycle $\pi_{2i-1}$ invariant and maps the other basic one-cycle $\pi_{2i}$ to $\pi_{2i-1} - \pi_{2i}$. As a consquence, the torus wrapping numbers $(n^i_a, m_a^i)$ of a one-cycle are mapped to $(n_a^i + m_a^i, - m_a^i)$. Regarding the orbifold-invariant bulk and exceptional basis of three-cycles, their behavior under the $\OR$ projection is captured in table~\ref{Tab:OR-on-Z2Z6p}. The orientifold projection maps\footnote{The transformation properties for the exceptional three-cycles are given here under the assumption that the $\OR$-plane is the exotic O6-plane. In general, an exceptional divisor is  multiplied with the factor $(-\eta_{\OR} \eta_{\OR \Z_2^{(k)}})$ under the orientifold projection.}  a $\Z_2^{(k)}$ exceptional divisor $e_{xy}^{(k)}$ to the exceptional divisor $e_{x' y'}^{(k)}$, where $x', y'$ denote the orientifold images of the fixed points $x, y$. Together with the transformation behavior of the one-cycle along $T_{(k)}^2$ the orbifold-invariant exceptional cycles transform as indicated in table~\ref{Tab:OR-on-Z2Z6p}.

\begin{SCtable}[50][h]
$\begin{array}{|c|c||c||c|}\hline
\muc{4}{|c|}{\text{\bf Orientifold projection on $T^6/(\Z_2 \times \Z_6' \times \OR)$ on {\bf AAA}}}
\\\hline\hline
\OR(\rho_1) & \OR(\rho_2) & \OR(\varepsilon_{\alpha}^{(k)}) & \OR(\tilde{\varepsilon}_{\alpha}^{(k)}) 
\\\hline\hline
\rho_1 & \rho_1 - \rho_2 &  \, \varepsilon_{\beta}^{(k)} &   \, \left(\tilde{\varepsilon}_{\beta}^{(k)} - \varepsilon_{\beta}^{(k)}\right) 
\\
& & \muc{2}{|c|}{\alpha=\beta=1,2,3 ; \; \alpha=4,5 \leftrightarrow \beta=5,4 
}
\\\hline
\end{array}$
\caption{Orientifold projection on bulk three-cycles $\rho_{i, i \in \{1,2\}}$ and exceptional three-cycles $\varepsilon_{\alpha,\alpha \in \{1 \ldots 5\}}$, $\tilde{\varepsilon}_{\alpha,\alpha \in \{1 \ldots 5\}}$
on $T^6/(\Z_2 \times \Z_6' \times \OR)$ with $\eta=-1 = \eta_{\OR}$ and {\bf AAA} lattice.}
\label{Tab:OR-on-Z2Z6p}
\end{SCtable}


\section{$USp(2N)$ and $SO(2N)$ gauge group enhancement}\label{S:USpbranes}
There are three distinct underlying motivations for classifying the $\OR$ invariant D6-branes with a $USp(2N)$ gauge group. In the first place, the D6-branes with $USp(2)$ factors can be used as probe branes in the K-theory constraints, which have to be satisfied in order for the model to be globally consistent. Secondly, in the next section, these D6-branes will be crucial to derive the sufficient conditions for the existence of discrete symmetries. And thirdly, a $USp(2)$ gauge group can serve as the $SU(2)_L$ gauge factor in MSSM model building, and similarly as the $SU(2)_R$ gauge group in left-right symmetric models. Here, the classification of $USp(2N)$ gauge group enhancements is reviewed for the {\bf AAA} lattice configuration with the $\OR$-plane as the exotic O6-plane ($\eta_{\OR} = -1$). For other configurations, the reader is kindly encouraged to consult~\cite{Honecker:2012qr}.  
    
The necessary condition for a D6-brane to carry a $USp(2N)$ gauge factors is formulated as the requirement that the fractional three-cycle $\Pi_a^{\text{frac}}$ and its orientifold image $\Pi_{a'}^{\text{frac}}$ belong to the same homology class. This requirement leads to a topological condition relating the discrete Wilson lines and displacements of the fractional three-cycle to the RR-charges of the O6-planes~\cite{Forste:2010gw}, see table~\ref{Tab:OR-inv-branes}. Due to the fact that all three two-tori are tilted (as a consequence of the $\Z_6'$ action), several combinations of  $(\vec{\tau_a})$ and $(\vec{\sigma_a})$ can lead to $SO(2N_a)$ enhancement, instead of $USp(2N_a)$ enhancement. In~\cite{Honecker:2012qr} it was argued that additional factors $(-)^{\sigma_a^i \tau_a^i}$ have to be inserted in the M\"obius amplitude per tilted two-torus $T_{(i)}^2$ on which the bulk one-cycle is parallel to its own orientifold image. Consequently, the M\"obius does not necessarily contribute negatively to the full beta-function coefficient, and gauge group enhancement of the form $U(N_a) \hookrightarrow SO(2N_a)$ can occur.  
\begin{table}[h]
\begin{center}
\begin{equation*}
\begin{array}{|c|c|}\hline
\muc{2}{|c|}{\text{\bf $\OR$ inv. three-cycles on $T^6/(\Z_2 \times \Z_{6}' \times \OR)$ ($\eta=-1$)}}
\\\hline\hline
a\pp \text{orbit} &  \text{\bf Topological condition}
\\\hline\hline  \rule{0pt}{4ex}
\OR & \left(\eta_{\OR\Z_2^{(1)}},\eta_{\OR\Z_2^{(2)}},\eta_{\OR\Z_2^{(3)}} \right) \stackrel{!}{=}  \left(  (-1)^{\delta^2_a + \delta^3_a} ,  (-1)^{\delta^1_a + \delta^3_a} ,  (-1)^{\delta^1_a + \delta^2_a} \right)
\\  \rule{0pt}{4ex}
\OR\Z_2^{(i)} & \left(\eta_{\OR\Z_2^{(i)}},\eta_{\OR\Z_2^{(j)}},\eta_{\OR\Z_2^{(k)}} \right) \stackrel{!}{=}  \left(  (-1)^{\delta^j_a + \delta^k_a} , - (-1)^{\delta^i_a + \delta^k_a} , - (-1)^{\delta^i_a + \delta^j_a} \right)   \rule{-1pt}{4ex}
\\
\hline
\end{array}
\end{equation*}
\end{center}
\caption{ 
Topological condition for a fractional D6-brane $a$ on $T^6/(\Z_2 \times \Z_{6}' \times \OR)$ with $\eta=-1$ to wrap an orientifold invariant three-cycle with $\delta^i_a \equiv  \sigma^i_a \tau^i_a \in \{0,1\}$~\cite{Forste:2010gw} for $\eta_{\OR}=-1$ (for other choices, replace $\eta_{\OR\Z_2^{(i)}} \to - \eta_{\OR} \,\eta_{\OR\Z_2^{(i)}}$ on the l.h.s.). 
Specific combinations of  $(\vec{\sigma}_a;\vec{\tau}_a)$ provide orientifold invariant D6-branes
parallel to each of the O6-plane orbits, leading to  $USp(2M_a)$ or  $SO(2M_a)$ gauge group enhancements.
}
\label{Tab:OR-inv-branes}
\end{table}

\noindent The $\OR$-invariant fractional three-cycles with enhanced gauge groups can be classified according to which O6-plane they are parallel to and depending on the choices of discrete Wilson lines $\vec{\tau_a}$ and diplacements $\vec{\sigma_a}$. The latter criterium determines whether the $U(N_a)$ gauge groups is enhanced to a $USp(2N_a)$ or a $SO(2 N_a)$ factor, while the first criterium determines the matter content in the antisymmetric ($\Anti_a$) and symmetric ($\Sym_a$) representation~\cite{Honecker:2012qr}:
\begin{itemize}
\item \framebox{All 108 D6-branes $\pp$ $\OR$ with $\sigma_a^i \tau_a^i  = 0, \, \forall\, i$: $USp(2N_a)$} with one chiral multiplet in the $(\Anti_a)$ representation from the $(\omega^k a) (\omega^k a)'_{k=1,2}$ sector
at relative angle $\pm \pi (-\frac{2}{3},\frac{1}{3},\frac{1}{3})$,
\item \framebox{All 4 D6-branes $\pp$ $\OR$ with $\sigma_a^i \tau_a^i  = 1, \, \forall\, i$: $SO(2N_a)$} with one chiral multiplet in the $(\Anti_a)$ representation from the $(\omega^k a) (\omega^k a)'_{k=1,2}$ sector,
\item \framebox{All $3 \times 36$ D6-branes $\pp$ $\OR\Z_2^{(i)}$ with $(\sigma_a^i \tau_a^i,  \sigma_a^j \tau_a^j ,  \sigma_a^k \tau_a^k ) =(1,0,0)$: $USp(2N_a)$} with one chiral multiplet in the $(\Sym_a)$ representation from the $(\omega^k a) (\omega^k a)'_{k=1,2}$ sector,
\item \framebox{All $3 \times 12$ D6-branes $\pp$ $\OR\Z_2^{(i)}$ with $(\sigma_a^i \tau_a^i,  \sigma_a^j \tau_a^j ,  \sigma_a^k \tau_a^k ) =(0,1,1)$: $SO(2N_a)$} with one chiral multiplet in the $(\Sym_a)$ representation from the $(\omega^k a) (\omega^k a)'_{k=1,2}$ sector.
\end{itemize}
\noindent The discrete $\Z_2^{(k)}$ eigenvalues $\tau_2^{\Z_2^{(k)}}$ do not play a r\^ole in the gauge group enhancement, but enter as an overall factor of four in the counting above. For a different choice of the exotic O6-plane, $USp(2N_a)$ and $SO(2N_a)$ gauge enhancement still occurs, but for different patterns of the discrete Wilson lines and displacements, as discussed in appendix B.1.1.~of \cite{Honecker:2012qr}. 

\section{Discrete symmetries}\label{S:DiscreteSymm}
The consistency of intersecting D6-brane models relies on global conditions such as the RR tadpole cancellation conditions and the K-theory constraints. The first ensure the absence of non-Abelian gauge anomalies, while the latter are necessary to cancel the D-brane charges not captured by the RR tadpole cancellation conditions. A third element guaranteeing consistency in string model building is the generalized Green-Schwarz mechanism, by which also the mixed Abelian/non-Abelian and purely Abelian gauge anomalies as well as mixed Abelian-gravitational anomalies vanish. In that process, the $U(1)_a$ inside the $U(N_a)$ gauge factor acquires a St\"uckelberg mass by eating an RR-axion, if the coupling to the St\"uckelberg term is non-zero. It might, however, happen that linear combinations $U(1)_X=\sum_a q_a U(1)_a$ arise as anomaly-free and massless gauge symmetries with vanishing St\"uckelberg-couplings. In terms of homology, the constraints on the existence of a massless $U(1)_X$ for toroidal orientifolds can be written as:
\begin{equation} \label{Eq:MasslessU1}  
 \Pi^{\text{even}}_i \circ \sum_a q_a \, N_a \, \Pi_a =0 \qquad \forall \,  i
 ,
\end{equation}
where $ \Pi^{\text{even}}_i$ represent the $\OR$-even three-cycles under the orientifold projection.

Generically, these constraints will not be fulfilled for arbitrary linear combinations, and the $U(1)_X$ will acquire a St\"uckelberg mass. In those cases, the $U(1)_X$ will no longer act as a local symmetry, but rather survive as a global perturbative symmetry. Non-perturbative effects like D-brane instantons break these global continuous symmetries further. However, the discrete $\Z_n$ subsymmetries of these continuous symmetries are not violated~\cite{BerasaluceGonzalez:2011wy} by non-perturbative effects, and serve as effective symmetries to constrain the form of the perturbative and non-perturbative superpotential. The conditions for the existence of discrete $\Z_n$ symmetries generalise the masslessness conditions in~(\ref{Eq:MasslessU1}):
\begin{equation}\label{Eq:Zn-condition}
 \Pi^{\text{even}}_i \circ \sum_a k_a \, N_a \, \Pi_a =0 \text{ mod } n \qquad \forall i
 ,
\end{equation}  
where the rational coefficients $q_a$ are replaced by integers $k_a$ satisfying $0 \leqslant k_a < n$ and \mbox{$\gcd(k_a,k_b \ldots n)=1$} for inequivalent discrete $\Z_n \subset \sum_a k_a U(1)_a$ symmetries.   

Zooming in on the toroidal orbifold $T^6/(\Z_2\times \Z_6' \times \OR)$ with discrete torsion and the $\OR$-plane as the exotic O6-plane ($\eta=-1 = \eta_{\OR}$), a basis for the $\OR$-even and $\OR$-odd three-cycles was given in~\cite{Honecker:2013hda}:
\begin{equation}\label{Eq:Z2Z6p_even+odd_basis_eta-1}
\begin{array}{l@{\hspace{0.4in}}l}
\Pi^{\text{even},\unity}_0 =\rho_1,
& \Pi^{\text{odd},\unity}_0 =-\rho_1 + 2 \, \rho_2 ,
\\
\Pi^{\text{even},\Z_2^{(k)}}_{\alpha \in \{1,2,3\}} = \varepsilon_{\alpha}^{(k)},
&\Pi^{\text{odd},\Z_2^{(k)}}_{\alpha \in \{1,2,3\}} = - \varepsilon_{\alpha}^{(k)}   + 2 \, \tilde{\varepsilon}_{\alpha}^{(k)},
\\
\Pi^{\text{even},\Z_2^{(k)}}_{4} = \varepsilon_4^{(k)} +  \varepsilon_5^{(k)},
&\Pi^{\text{odd},\Z_2^{(k)}}_{4} = 2 \, (\tilde{\varepsilon}_4^{(k)} + \tilde{\varepsilon}_5^{(k)})  - ( \varepsilon_4^{(k)} +  \varepsilon_5^{(k)}),
\\
\Pi^{\text{even},\Z_2^{(k)}}_{5} = 2 \, (\tilde{\varepsilon}_4^{(k)} - \tilde{\varepsilon}_5^{(k)})  - (\varepsilon_4^{(k)} -  \varepsilon_5^{(k)}),
&\Pi^{\text{odd},\Z_2^{(k)}}_{5} = \varepsilon_4^{(k)} -  \varepsilon_5^{(k)}.
\end{array}
\end{equation}
From the intersection numbers for the bulk and exceptional three-cycles in equations~(\ref{Eq:IntersectionBulk}) and (\ref{Eq:IntersectionExceptional}), respectively, the intersection form for the $\OR$-even and $\OR$-odd three-cycle basis then reads:
\begin{equation}
\Pi^{\text{even},\Z_2^{(k)}}_{\tilde{\alpha}} \circ \Pi^{\text{odd},\Z_2^{(l)}}_{\tilde{\beta}} = \delta^{kl} \delta_{\tilde{\alpha}\tilde{\beta}} \times 
\left\{\begin{array}{cr}
8 & \tilde{\alpha}=0
\\ - 8 & 1 \ldots 3
\\ - 16 & 4
\\ 16 & 5
\end{array} \right. \, 
\qquad \text{ with } \quad 
\Z_2^{(0)} \equiv \unity
,
\end{equation}
indicating that the basis is not a unimodular basis, as a consequence of the tilted nature of all three two-tori. Inserting the sixteen basic $\OR$-even three-cycles of equation~(\ref{Eq:Z2Z6p_even+odd_basis_eta-1}) into the conditions for the existence of discrete $\Z_n$ symmetries in equation~(\ref{Eq:Zn-condition}) leads to the sixteen conditions on the lefthand side of equation~(\ref{Eq:Z2Z6p-necessary+sufficient_discrete}):

\begin{equation}\label{Eq:Z2Z6p-necessary+sufficient_discrete}
{\footnotesize
\hspace*{-0.4in}\sum_a k_a N_a \; \left(\begin{array}{c}  Y_a \\\hline -y_{a,1}^{(1)}  \\ -y_{a,2}^{(1)}  \\ -y_{a,3}^{(1)} \\  - (y_{a,4}^{(1)} + y_{a,5}^{(1)})\\ 2 \, (x_{a,4}^{(1)} - x_{a,5}^{(1)}) + (y_{a,4}^{(1)} - y_{a,5}^{(1)})
\\\hline -y_{a,1}^{(2)}  \\ -y_{a,2}^{(2)}  \\ -y_{a,3}^{(2)} \\  - (y_{a,4}^{(2)} + y_{a,5}^{(2)})\\ 2 \, (x_{a,4}^{(2)} - x_{a,5}^{(2)}) + (y_{a,4}^{(2)} - y_{a,5}^{(2)})
\\\hline -y_{a,1}^{(3)}  \\ -y_{a,2}^{(3)}  \\ -y_{a,3}^{(3)} \\  - (y_{a,4}^{(3)} + y_{a,5}^{(3)})\\ 2 \, (x_{a,4}^{(3)} - x_{a,5}^{(3)}) + (y_{a,4}^{(3)} - y_{a,5}^{(3)})
\end{array}\right)}
\stackrel{!}{=} 0 \text{ mod } n 
  \stackrel{!}{=} 
{\footnotesize \sum_a k_a N_a \; \left(\begin{array}{c}  \frac{Y_a - \sum_{i=1}^3 [ y^{(i)}_{a,1} +  y^{(i)}_{a,2} + y^{(i)}_{a,3} ]}{4} \\ \frac{Y_a - [ y^{(1)}_{a,1} +  y^{(1)}_{a,2} + y^{(1)}_{a,3} ]}{2} \\ 
\frac{Y_a - [ y^{(2)}_{a,1} +  y^{(2)}_{a,2} + y^{(2)}_{a,3} ]}{2} \\ - \frac{   y^{(2)}_{a,1} +y^{(2)}_{a,3}+ y^{(3)}_{a,1} +y^{(3)}_{a,3}  }{2} \\ - \frac{y^{(1)}_{a,1} + y^{(1)}_{a,3} + y^{(3)}_{a,2} +y^{(3)}_{a,3} }{2} \\ - \frac{y^{(1)}_{a,2} + y^{(1)}_{a,3} + y^{(2)}_{a,2} +y^{(2)}_{a,3}}{2} 
\\\hline
\frac{Y_a + [y^{(1)}_{a,3} + x^{(1)}_{a,4} + y^{(1)}_{a,4} - x^{(1)}_{a,5}] + \sum_{j=2}^3 [y^{(j)}_{a,2} - (y^{(j)}_{a,4} + y^{(j)}_{a,5})]}{4}
\\
\frac{Y_a + \sum_{j=1,2} [y^{(j)}_{a,1} - x^{(j)}_{a,4}  + x^{(j)}_{a,5} + y^{(j)}_{a,5}] + [y^{(3)}_{a,3} + x^{(3)}_{a,4} + y^{(3)}_{a,4} - x^{(3)}_{a,5}] }{4}
\\
\frac{Y_a + [y^{(2)}_{a,2} - (y^{(2)}_{a,4} + y^{(2)}_{a,5})]}{2}
\\
\frac{Y_a + [y^{(1)}_{a,1} - x^{(1)}_{a,4}  + x^{(1)}_{a,5} + y^{(1)}_{a,5}]}{2}
\\
- \frac{y^{(2)}_{a,4} + y^{(2)}_{a,5} +  y^{(3)}_{a,4} + y^{(3)}_{a,5} }{2}
\\
-\frac{ x_{a,4}^{(1)} - x^{(1)}_{a,5}  + y^{(1)}_{a,5}+ x_{a,4}^{(2)} - x^{(2)}_{a,5} + y^{(2)}_{a,5} }{2}
 \\\hline
 \frac{Y_a + \sum_{i=1}^3 [y^{(i)}_{a,3} + x^{(i)}_{a,4} + y^{(i)}_{a,4} - x^{(i)}_{a,5}] }{4} \\ \frac{Y_a + y^{(1)}_{a,3}  + x^{(1)}_{a,4} + y^{(1)}_{a,4} - x^{(1)}_{a,5}}{2} \\
\frac{Y_a + y^{(2)}_{a,3}  + x^{(2)}_{a,4} + y^{(2)}_{a,4} - x^{(2)}_{a,5} }{2} \\ \frac{Y_a + \sum_{i=1}^3 y^{(i)}_{a,3}}{2}
\end{array}\right)
.
}
\end{equation}

\noindent However, as the basic three-cycles in equation~(\ref{Eq:Z2Z6p_even+odd_basis_eta-1}) do not form a unimodular basis but only a sublattice, additional conditions are expected to arise from shorter $\OR$-even three-cycles, which 
are fractional linear combinations of the basis in equation~(\ref{Eq:Z2Z6p_even+odd_basis_eta-1}). In order to find the second set of conditions on the righthand side of equation~(\ref{Eq:Z2Z6p-necessary+sufficient_discrete}), we turn to the  fractional three-cycles in section~\ref{S:USpbranes} with enhanced gauge factors. These three-cycles are parallel to one of the O6-planes and are inherently even under the orientifold projection. As an example, the fractional three-cycles parallel to the $\OR$-plane with discrete displacements $(\vec{\sigma}) = (0,1,1)$ and $(\vec{\sigma}) = (1,1,0)$ are explicitly expanded in terms of the basic $\OR$-even cycles of equation~(\ref{Eq:Z2Z6p_even+odd_basis_eta-1}):
\begin{equation}\label{Eq:ExampleORinvcycle}
\begin{aligned}
\Pi^{\text{frac}, \OR}_{(\sigma^1,\sigma^2,\sigma^3) = (0,1,1)}  \stackrel{\tau^2 = 0 = \tau^3}{=} &\frac{\Pi_0^{\text{even}, \1}}{4} + \frac{(-)^{\tau^{\Z_2^{(1)}}}}{4} \left( - \Pi^{\text{even},\Z_2^{(1)}}_3 + \frac{- \Pi^{\text{even},\Z_2^{(1)}}_4+ \Pi^{\text{even},\Z_2^{(1)}}_5}{2}   \right)\\
& + \sum_{j=2,3} \frac{(-)^{\tau^{\Z_2^{(j)}}}}{4} \left( - \Pi^{\text{even},\Z_2^{(j)}}_2 + (-)^{\tau^1} \, \Pi^{\text{even},\Z_2^{(j)}}_4  \right),\\
\Pi^{\text{frac}, \OR}_{(\sigma^1,\sigma^2,\sigma^3) = (1,1,0)}  \stackrel{\tau^1 = 0 = \tau^2}{=} & \frac{\Pi_0^{\text{even}, \1}}{4} + \sum_{i=1,2}  \frac{(-)^{\tau^{\Z_2^{(i)}}}}{4}
\left( - \Pi^{\text{even},\Z_2^{(i)}}_1 - (-)^{\tau^3} \, \frac{ \Pi^{\text{even},\Z_2^{(i)}}_4+ \Pi^{\text{even},\Z_2^{(i)}}_5}{2}   \right)\\
& +  \frac{(-)^{\tau^{\Z_2^{(3)}}}}{4} \left( - \Pi^{\text{even},\Z_2^{(3)}}_3 + \frac{- \Pi^{\text{even},\Z_2^{(3)}}_4+ \Pi^{\text{even},\Z_2^{(3)}}_5}{2}   \right).
\end{aligned}
\end{equation}
Comparing the structure of both expansions tells us that the two three-cycles will yield different conditions when inserting them into equation~(\ref{Eq:Zn-condition}). However, not all new conditions derived from the shortest $\OR$-invariant fractional three-cycles in this way will be independent. Given the conditions on the lefthand side in equation~(\ref{Eq:Z2Z6p-necessary+sufficient_discrete}) and the property that some of the fractional three-cycles with enhanced gauge group can be written as linear combinations of others, there are only sixteen other conditions that have to be taken into account.

The second (central) block in the set of conditions on the righthand side of~(\ref{Eq:Z2Z6p-necessary+sufficient_discrete})
are obtained from the three-cycles in equation~(\ref{Eq:ExampleORinvcycle}) by taking particular values for the $\Z_2$ eigenvalues, or by taking linear combinations with different $\Z_2$ eigenvalues or discrete Wilson lines: the first two rows follow from the fractional three-cycles with $\Z_2$ eigenvalues $(+++)$, the next two rows follow by adding the $\OR$-invariant three-cycles in~(\ref{Eq:ExampleORinvcycle}) with different $\Z_2$ eigenvalues, and the last two rows result from subtracting the first three-cycle with $\tau^1 = 0$ from the first three-cycle with $\tau^1=1$, and subtracting the second three-cycle with $\tau^3 = 0$ from the second three-cycle with $\tau^3 = 1$. 

In total, there are 32 conditions for the existence of discrete $\Z_n$ symmetries on the orientifold  $T^6/(\Z_2\times \Z_6' \times \OR)$: sixteen {\it necessary} conditions on the lefthand side of equation~(\ref{Eq:Z2Z6p-necessary+sufficient_discrete}) and sixteen {\it sufficient} conditions on the righthand side of equation~(\ref{Eq:Z2Z6p-necessary+sufficient_discrete}). Nonetheless, not all conditions will be inequivalent or independent, as the space of $\OR$-even three-cycles is a $h_{21} +1 = 16$ dimensional vector space. Hence, at most sixteen relations can serve as independent constraints, and in practice many of the conditions in~(\ref{Eq:Z2Z6p-necessary+sufficient_discrete}) will turn out to be trivially satisfied, identical or related to each other by linear combinations.

\section{Model building and a global six-stack Pati-Salam model}\label{S:GlobalPatiSalam}

At this point, we have collected all the necessary ingredients to review model building efforts on the orbifold  $T^6/(\Z_2\times \Z_6' \times \OR)$ with discrete torsion. The three $\Z_2$ twisted sectors force a fractional three-cycle to be stuck at the $\Z_2$ fixed points on all three two-tori, projecting out the chiral multiplets in the adjoint representation that allowi for a continuous displacement of the D-branes along the two-tori, the so-called `position moduli'. However, this property does not guarantee the complete absence of chiral multiplets in the adjoint representation. From the open string sectors $a (\omega^k a)_{k=1,2}$ of intersecting $\Z_6'$ orbifold images,
additional chiral fields in the adjoint representation can arise, encoding the `deformation moduli' of the intersection point. In~\cite{Honecker:2012qr} it was shown how the presence of this additional adjoint matter depends on the values of the discrete displacements $(\vec{\sigma})$ and discrete Wilson lines $(\vec{\tau})$. Combining the bulk RR tadpole conditions with the requirement of a completely rigid three-cycle for the QCD stack points towards the use of the shortest fractional three-cycles for model building. 

The particular chiral structure of the Standard Model spectrum puts further restrictions on the types of globally consistent models that can be found on this orbifold. Finding global
three-generation MSSM or left-right symmetric models with the QCD stack wrapped on rigid three-cycles is obstructed by the non-vanishing of the RR tadpoles in the $\Z_2^{(k)}$ twisted sectors. With regard to global Pati-Salam models, two types of models were found: five-stack D6-brane models with gauge group $U(4)_a\times U(2)_L\times U(2)_R \times U(2)_d \times U(2)_e$ and six-stack D6-brane models with gauge group $U(4)_a \times U(2)_L \times U(2)_R \times U(4)_d \times U(2)_e \times U(2)_f$. The global five-stack models were discussed in detail in~\cite{Honecker:2012qr}. Here, we will discuss some of the phenomenological properties of the global six-stack models, such as the chiral and non-chiral spectrum, Yukawa- and other three-point couplings, and the discrete $\Z_n$ symmetries arising from the full gauge group. 

An exemplary D6-brane configuration can be found in table 45 of~\cite{Honecker:2012qr}, for which the chiral spectrum is given here in table~\ref{Tab:OtherDecentPatiSalamC} and the non-chiral spectrum in table~\ref{Tab:OtherDecentPatiSalamNC}. A first observation from the chiral spectrum is that the three left-handed and right-handed generations of quarks and leptons come from different sectors: the $ab$ ($ac'$) intersections produce two left(right)-handed generations, while a third left(right)-handed generation arises from the $ab'$ ($ac$) intersections. The Higgs-sector comprises two chiral states from the $bc'$ sector and one non-chiral pair from the $bc$ sector (`chirality' here refers to the $U(1) \subset U(2)$ gauge factors). Several chiral exotic matter states ensure the vanishing of non-Abelian gauge anomalies in the global model, but none of these exotics is charged under the QCD gauge group. In the non-chiral spectrum, additional exotic matter arises in the bifundamental representation of the QCD stack and one of the hidden gauge groups. Furthermore, both $U(4)$ gauge groups come with two non-chiral matter pairs in the antisymmetric representation.

\begin{table}[h]
\begin{center}
\hspace*{-0.3in}\begin{tabular}{|c||c|c|c|c||c|c|c|c|c|c|c|}
\hline \multicolumn{5}{|c||}{\bf Chiral spectrum of a 6-stack Pati-Salam model}& \multicolumn{7}{|c|}{\bf Discrete charges}\\
\hline \hline
Matter & Sector & {\scriptsize $U(4)_a\times U(2)^2  \times U(4)_d \times U(2)^2 $}& {\scriptsize $(Q_a, Q_b, Q_c, Q_d, Q_e, Q_f)$}&$Q_Z$& \multicolumn{2}{|c|}{$\Z_4^a$}&$\Z_2^b$&$\Z_2^c$&$\Z_4^d$&$\Z_2^e$&$\Z_2^f$ \\
\hline  $Q_{ab}^{(i)}$&$ab$&$2 \times ({\bf 4}, {\bf\ov 2}, \1,\1,\1,\1)$ &(1,-1,0,0,0,0) &$-1$&1&0&0&0&0&0&0\\
$Q_{ab'}$&$ab'$&  $({\bf4}, {\bf 2}, \1,\1,\1,\1)$&(1,1,0,0,0,0)&$1$&1&2&0&0&0&0&0\\
$U_{ac}$ & $ac$ & $({\bf\ov 4}, \1,{\bf 2},\1,\1,\1)$ &(-1,0,1,0,0,0)&$-1$&3&2&1&1&0&0&0 \\
$U_{ac'}^{(i)}$& $ac'$ &  $2 \times ({\bf\ov 4}, \1,{\bf\ov 2},\1,\1,\1)$&(-1,0,-1,0,0,0) &$1$&3&0&1&1&0&0&0\\
$h^{(i)}_{bc'}$&$bc'$&$2 \times (\1, {\bf2}, {\bf 2},\1,\1,\1)$&(0,1,1,0,0,0) &$0$&0&0&1&1&0&0&0\\
$X_{bd}$& $bd$ & $(\1, {\bf 2}, \1, {\bf \ov 4},\1,\1) $&(0,1,0,-1,0,0)&$1$&0&1&0&0&3&0&0\\
$X_{bd'}$&$bd'$&  $(\1, {\bf \ov  2}, \1, {\bf \ov 4},\1,\1) $ &(0,-1,0,-1,0,0)&$-1$&0&3&0&0&3&0&0\\
$X_{bf}$&$bf$&$(\1, {\bf 2}, \1, \1,\1, {\bf \bar 2}) $&(0,1,0,0,0,-1)&$0$&0&0&1&0&0&0&1\\
$X_{bf'}$&$bf'$&$(\1, {\bf \ov  2}, \1, \1, \1, {\bf \ov 2}) $&(0,-1,0,0,0,-1)&$-2$&0&2&1&0&0&0&1\\
$X_{cd}$&$cd$& $(\1, \1, {\bf \ov 2}, {\bf 4},\1,\1) $   &(0,0,-1,1,0,0)&$1$&0&1&1&1&1&0&0 \\
$X_{cd'}$&$cd'$&  $(\1, \1, {\bf  2}, {\bf 4},\1,\1) $ & (0,0,1,1,0,0)&$-1$&0&3&1&1&1&0&0\\
$X_{cf}$&$cf$&  $(\1, \1, {\bf 2}, \1, \1, {\bf \ov 2}) $ & (0,0,1,0,0,-1)&$-2$&0&2&0&1&0&0&1\\
$X_{cf'}$&$cf'$&  $(\1, \1, {\bf \ov 2}, \1, \1, {\bf \ov 2}) $ & (0,0,-1,0,0,-1)&$0$&0&0&0&1&0&0&1\\
\hline
\end{tabular}
\caption{Chiral spectrum for a global six-stack Pati-Salam model with gauge group $U(4)_a \times U(2)_L \times U(2)_R \times U(4)_d \times U(2)_e\times U(2)_f$. The notation $Q_{xy}^{(i)}$ ($U^{(i)}_{xy}$) is introduced to denote the left(right)-handed quarks and leptons, while $h^{(i)}_{bc'}$ denotes the Higgs-doublets $\left(H_d, H_u\right)$ in the $bc'$ sector. Exotic chiral matter is denoted by $X_{xy}$. The superscript $i$ specifies the multiplicity of the state and is omitted when the respective state only occurs once. The columns on the righthand side list the charges of the various states under the discrete $\Z_n$ symmetries. The discrete charges of the left-handed quarks and leptons from the $ab$ sector can be set to zero by a $Q_Z$-shift, leading to the second column under the $\Z_4^a$ symmetry. \label{Tab:OtherDecentPatiSalamC}}
\end{center}
\end{table}

\subsection{Discrete symmetries}
In total, there are six Abelian gauge factors, one for each D6-brane stack. Four $U(1)$'s  combine into a massless and generation-dependent gauged  $U(1)_Z$ with conserved charge:
\begin{equation}
Q_Z = Q_b - Q_c + Q_e + Q_f.
\end{equation}
The five orthogonal directions acquire St\"uckelberg masses, as described in section~\ref{S:DiscreteSymm}, and survive as global perturbative $U(1)$ symmetries. Hence, five discrete $\Z_n$ symmetries are expected to arise from these global $U(1)$ symmetries that constrain not only the perturbative but also the non-perturbative terms in the superpotential. In~\cite{Honecker:2013hda}, the necessary and sufficient conditions from equation~(\ref{Eq:Z2Z6p-necessary+sufficient_discrete}) were explicitly written down and solved in order to find the independent $\Z_n$ symmetries for this global model. Here, the discrete symmetries and the discrete charges for the chiral as well as for the non-chiral states are listed in tables~\ref{Tab:OtherDecentPatiSalamC}  and~\ref{Tab:OtherDecentPatiSalamNC}, respectively. The inequivalent $\Z_n$ symmetries turn out to be the symmetries arising from a single gauge factor: $\Z_4^{a,d} \subset U(1)_{a,d}$ and $\Z_2^{b,c,e,f} \subset U(1)_{b,c,e,f}$. The $\Z_2$ symmetries from $U(1)_{b,c}$ both have the effect of a matter parity and only differ in their action on the chiral exotic matter. The discrete $\Z_2$ and $\Z_4$ symmetries arising from the hidden gauge groups act trivially on the quarks and leptons and Higgses, but act non-trivially on the exotic matter. The $\Z_2^e$ symmetry is of particular interest, as it represents a discrete symmetry which does not act on the chiral matter at all, yet acts non-trivially on the non-chiral states arising at intersections with the $e$ stack. Finally, the $\Z_4^a$ symmetry acts differently on the left-and rigt-handed quarks and leptons, and becomes a generation-dependent discrete symmetry after a $Q_Z$-shift setting the charges of the $ab$ states to zero. Although we have listed six discrete symmetries, only five of them are truly independent: the two $\Z_4$ symmetries and three of the $\Z_2$ symmetries. The fourth $\Z_2$ symmetry is related to the other three discrete $\Z_2$ symmetries by virtue of the gauged $U(1)_Z$ symmetry.

\begin{table}[h]
\begin{center}
\hspace*{-0.2in}\begin{tabular}{|c|c|c|c||c|c|c|c|c|c|c|}
\hline \multicolumn{4}{|c||}{\bf Non-chiral spectrum  of a 6-stack Pati-Salam model}& \multicolumn{7}{|c|}{\bf Discrete charges}\\
\hline \hline sector  & {\scriptsize $U(4)_a \times U(2)^2 \times U(4)_d \times U(2)^2 $}& {\scriptsize $(Q_a, Q_b, Q_c, Q_d, Q_e,Q_f)$} &$Q_Z$& \multicolumn{2}{|c|}{$\Z_4^a$}&$\Z_2^b$&$\Z_2^c$&$\Z_4^d$&$\Z_2^e$&$\Z_2^f$\\
\hline
$aa'$& $2 \times [({\bf {6}_\Anti}, \1,\1,\1,\1,\1) + h.c.]$ & ($\pm 2$,0,0,0,0,0)&$0$&2 &2&0&0&0&0&0 \\
$dd'$& $2 \times [(\1,\1,\1,{\bf {6}_\Anti}, \1,\1) + h.c.]$ & (0,0,0,$\pm 2$,0,0)&$0$&0 &0&0&0&2&0&0  \\
 $ee'$ & $2 \times [(\1,\1,\1, \1, {\bf 1_{\Anti}},\1) + h.c.]$ & (0,0,0,0,$\pm 2$,0)&$\pm2$&0 &2&0&0&0&0&0 \\
 $ff'$ &  $2 \times [(\1,\1,\1, \1, \1,{\bf 1_{\Anti}}) + h.c.]$  & (0,0,0,0,0,$\pm 2$)&$\pm2$&0 &2&0&0&0&0&0 \\
 $ad$ &   $ ({\bf 4}, \1,\1, {\bf \bar 4},\1,\1) + h.c.$ & $(\pm 1, 0, 0, \mp 1, 0 , 0)$&$0$&1 &1&0&0&3&0&0  \\
  $ae'$ &  $ ({\bf 4}, \1,\1, \1, {\bf  2},\1) + h.c.$ &$(\pm 1, 0, 0, 0, \pm 1, 0)$&$\pm1$&1 &2&0&0&0&1&0 \\
 $bc$ & $ (\1, {\bf 2}, {\bf \bar 2},\1,\1,\1) + h.c. $  &$(0, \pm 1, \mp 1, 0, 0, 0)$ &$\pm2$&0 &2&1&1&0&0&0 \\
 $de$ &  $(\1, \1, \1, {\bf 4}, {\bf \bar 2},\1) + h.c. $  &$(0, 0, 0, \pm 1, \mp 1 , 0)$&$\mp1$&0 &3&0&0&1&1&0  \\
 $df'$ &  $(\1,\1,\1, {\bf 4},\1, {\bf 2}) + h. c. $&$(0, 0, 0, \pm 1, 0, \pm 1)$&$\pm 1$& 0&1&0&0&1&0&1 \\
 $ef$ & $ (\1,\1,\1,\1, {\bf 2}, {\bf \bar 2}) + h.c.  $&$(0, 0, 0, 0,\pm 1, \mp 1)$&$0$&0 &0&0&0&0&1&1 \\
 \hline
\end{tabular}
\caption{Non-chiral spectrum for a global six-stack Pati-Salam model with gauge group $U(4)\times U(2)_L \times U(2)_R \times U(4)_d \times U(2)_e \times U(2)_f$. The sector $bc$ contains one non-chiral pair of Higgs doublets $(H_{bc},\ov{H}_{bc})$
in the $(\1, {\bf 2}, {\bf \bar 2},\1,\1,\1) + h.c. $ representation.
The discrete charges under the $\Z_n$ symmetries are only listed for one state per non-chiral pair in the columns on the righthand side. The second column under the $\Z_4^a$ symmetry results from a $Q_Z$-shift of the first column. \label{Tab:OtherDecentPatiSalamNC}}
\end{center}
\end{table}

\subsection{The superpotential}

As a last step in the analysis, some words are devoted to the superpotential of this six-stack Pati-Salam model. The superpotential generically consists of a perturbative part, including the Yukawa and three-point couplings arising from worldsheet instantons, and a non-perturbative part containing couplings arising from D-brane instanton corrections. 

Regarding the Yukawa and other three-point couplings, their existence depends on the ability to construct a closed sequence $[x,y,z]$ outlining a closed triangle with edges $[x,y], [y,z]$ and $[z,x]$ on the underlying torus, where $x, y$ and $z$ represent three intersecting D6-branes. The intersection points between the D6-branes serve as apexes for the triangle, at which the respective chiral states in the three-point coupling are localized. The coefficient of the three-point coupling is then exponentially suppressed by the triangular area enclosed by the three intersecting D6-branes in units of the inverse string tension $\alpha'$. If the three cycles intersect in a single point, the area of the triangle is zero and the coupling between the chiral states is of order ${\cal O}(1)$. 

Table~\ref{Tab:PS2Yukawa} lists a series of Yukawa and three-point couplings for the six-stack Pati-Salam model, together with the appropriate closed sequence, the enclosed triangle and its area. Various other three-point couplings pass the charge selection rule, i.e.~the coupling represents a gauge invariant operator under the full gauge group, but fail the closed sequence selection rule, as matter states are localized at intersection points of various orbifold images which do not allow to construct a closed triangle. For those three-point couplings, the coefficients merely vanish. Concerning the couplings in table~\ref{Tab:PS2Yukawa}, all of them are of order ${\cal O}(1)$, and all generations of quarks and leptons enter in one way or another in a Yukawa coupling. The non-chiral Higgses  $H_{bc},\ov{H}_{bc})$ from the $bc$ sector can be used to give mass to some of the chiral exotic states.

Exotic matter states, which do not acquire mass through the perturbative couplings, might acquire mass through higher order couplings or non-perturbative effects, such as D-brane instanton corrections~\cite{Blumenhagen:2009qh}. As discussed above, the geometry of this orbifold allows for the existence of rigid three-cycles, on which E2-instantons (Euclidean D2-branes) can wrap. In order for the E2-instanton to induce additional couplings for the exotic matter states, particular constraints on the intersection numbers between the E2-instanton and the D6-branes have to be satisfied. At this point also discrete symmetries come into play, as they are left unbroken by E2-instantons and therefore constrain the possible appearance of additional couplings for the exotic matter states.

\begin{table}[h]
\begin{center}
\begin{tabular}{|c|c|c|c|}
\hline \multicolumn{4}{|c|}{\bf Yukawa and other three-point couplings for a six-stack Pati-Salam Model} \\
\hline \hline \bf sequence  $[x,y,z]$ & \bf coupling & \bf triangle & \bf enclosed area\\
 \hline \hline  \rule{0pt}{3ex}     $[a,b,c]$ & $Q^{(3)}_{ab}\, H_{bc}\, U_{ac}$ & $\{6, \|=[5,6], 1\}$ & 0 \\
 \hline  \rule{0pt}{3ex}  $[a,b',c']$ & $Q_{ab'}\, \ov{H}_{bc} \, U^{(3)}_{ac'}$& $\{5,\|=[5,6],1\}$ & 0\\
 \hline \rule{0pt}{3ex}    $[a,b,(\omega c)']$ & $Q^{(2)}_{ab}\, h^{(1)}_{bc'}\, U^{(3)}_{ac'} $ &$\{6,5,1\}$ &0\\
   \rule{0pt}{3ex}  $[a, (\omega b),c']$ & $Q^{(3)}_{ab}\, h^{(1)}_{bc'}\, U^{(2)}_{ac'} $ &$\{5,6,1\}$ &0\\
  \rule{0pt}{3ex}   $[a, (\omega b),(\omega c)']$ & $Q^{(3)}_{ab}\, h^{(2)}_{bc'}\, U^{(3)}_{ac'} $ &$\{5,6,1\}$ &0\\
\hline  \rule{0pt}{3ex}   $[b,c,d]$ & $\ov{H}_{bc} \, X_{bd}\, X_{cd}$  & $\{6, \|=[5,6], 6\}$ & 0 \\
   \rule{0pt}{3ex} $[b,c,d']$ & $H_{bc}\, X_{bd'}\, X_{cd'}$  & $\{6, \|=[5,6], 6\}$ & 0 \\
\hline
\end{tabular}
\caption{Overview of the surviving Yukawa and three-point couplings for the global six-stack Pati-Salam model. The symbol $\|=[5,6]$ represents three parallel one-cycles going through the $\Z_2$ fixed points $5$ and $6$. The symbols $H_{bc}$ and $\ov{H}_{bc}$ are introduced to denote the non-chiral Higgs-doublet pairs $(H_u, H_d)$ arising at the $bc$ intersection.\label{Tab:PS2Yukawa}}
\end{center}
\end{table}

\section{Conclusions}\label{S:Conclusions}
We reviewed various geometric properties of rigid three-cycles on the orbifold $T^6/(\Z_2\times \Z_6' \times \OR)$ with discrete torsion, which allow to construct globally consistent models without matter multiplets in the adjoint representation. At first sight, the factorisable orbifold allows for four different lattice configurations depending on the orientation of the two-tori with respect to the $\OR$-invariant plane, but via a pairwise identification they can be brought back to two inequivalent lattices: {\bf AAA} and {\bf BBB}. For model building purposes, only the first configuration offers promising construction opportunities, which is illustrated with a global six-stack Pati-Salam model. The other configuration does not allow for the construction of global models with large hidden gauge groups.

The tilted nature of the three two-tori complicates the process of writing down the conditions for the existence of discrete $\Z_n$ symmetries on this orbifold. We present a method to find all inequivalent conditions, and we discuss the discrete $\Z_n$ symmetries for the constructed global six-stack Pati-Salam model. The analysis of the global model is concluded with an extended study of the
perturbative Yukawa and other three-point couplings in the superpotential.


\end{document}